\documentclass[referee]{aa}
\usepackage[latin1]{inputenc}
\usepackage[T1]{fontenc}
\usepackage{multirow}
\usepackage{graphics}
\usepackage{afterpage}
\usepackage{graphicx}
\usepackage[centerlast,it,small]{subfigure}
\usepackage{lscape}
\usepackage{longtable}
\usepackage{supertabular}

\begin{document}
\authorrunning{S. Lauger, D. Burgarella, and V. Buat}
\titlerunning{Spectro-morphology of galaxies}
\title{Spectro-Morphology of Galaxies : a multi-wavelength (UV-R) evolutionary method}
\author{S. Lauger, D. Burgarella \and V. Buat}
\institute{Laboratoire d'Astrophysique de Marseille, Traverse du Siphon B.P.8, F-13376~Marseille~Cedex~12}
\offprints{S. lauger, sebastien.lauger@oamp.fr}
\date{Received ... / Accepted ...}



\abstract{We present a quantitative method to classify galaxies, based on multi-wavelength data and elaborated from the properties of nearby galaxies. Our objective is to define an evolutionary method that can be used for low and high redshift objects. We estimate the concentration of light ($C$) at the galaxy center and the 180$^o$-rotational asymmetry ($A$), computed at several wavelengths, from ultraviolet (UV) to I-band. The variation of the indices of concentration and asymmetry with the wavelength reflects the proportion and the distribution of young and old stellar populations in galaxies. In general $C$ is found to decrease from optical to UV, and $A$ is found to increase from optical to UV: the patchy appearance of galaxies in UV with no bulge is often very different from their counterpart at optical wavelengths, with prominent bulges and more regular disks. The variation of $C$ and $A$ with the wavelength is quantified.  By this way, we are able to distinguish five types of galaxies that we call spectro-morphological types: compact, ringed, spiral, irregular and central-starburst galaxies, which can be differentiated by the repartition of their stellar populations. We discuss in detail the morphology of galaxies of the sample, and describe the morphological characteristics of each spectro-morphological type. We apply spectro-morphology to three objects at a redshift $z\sim1$ in the Hubble Deep Field North, that gives encouraging results for applications to large samples of high-redshift galaxies. This method of morphological classification could be used to study the evolution of the morphology with the redshift and is expected to bring observational constraints on scenarios of galaxy evolution.
}
\maketitle
\section{Introduction}
The morphology of galaxies is the result of physical processes that acted on these systems since their formation. In an attempt to organize galaxies and to understand their evolution, the Hubble-Sandage classification, based on the bulge-to-disk ratio, was built from the study of a sample of nearby bright galaxies (\cite{sandage61}). However, it now appears that this classification is not appropriate for faint nearby galaxies or small-sized objects. Moreover, with telescopes reaching out to redshifts $z\ge1$ the rate of unclassified galaxies seems to increase with the distance, and an important fraction has unusual morphology (\cite{brinchmann98}, Abraham et al. 1996, van den Bergh 2002). Theoretical models predict this evolution by accounting for an evolution of the galaxy morphology (\cite{baugh}). The origin of this morphological change with $z$ is likely twofold: the morphological K-correction (Burgarella et al. 2001) and the intrinsic galaxy evolution (\cite{giavalisco}). A first step in elaborating a method of morphological classification is to be able to classify galaxies  within morphological boxes  and to calibrate the method where the results can be checked by eye, \textit{i.e.} in the nearby universe. We also want to discriminate between merging and irregular galaxies, in order to confirm the validity of galaxy evolution models. The Hubble classification fails on this point (\cite{naim97}).
By comparing morphological characteristics of local and distant samples, it might be possible to constrain the mechanisms of formation and evolution of galaxies in the early universe. As a result, it is necessary to elaborate a new system of classification, which would be more objective than the Hubble classification. A scheme based on quantitative indices could have the advantage to not implicate human judgment and to be automatisable, in order to classify quickly large samples of galaxies. The quantification also allows us to measure the variation of the morphology with the wavelength ($\lambda$). Finally, in order to compare low and high-redshift galaxies, these parameters have to be robust against a degradation of spatial resolution and low signal-to-noise ratios ($S/N$).\\
For several years, a number of parameters have been proposed. Morgan (1958, 1959) originally used the index of light concentration in the center of galaxy - the basis of the Hubble scheme with the morphology of the spiral structure. \cite{kent} and \cite{abraham94} developed two quantitative measures of concentration. Abraham et al. (1996) also used an asymmetry index of galaxies developed  by \cite{schade95}. The resulting asymmetry-concentration diagram calibrated from a local sample allowed them to distinguish three morphological types in visible: E/SO galaxies, spiral galaxies and irregular/peculiar galaxies (Abraham et al. 1996; Bershady et al. 2000). Moreover, it seems possible to discriminate irregular galaxies from mergers by using a correlation between color index and asymmetry (Conselice et al. 2000).\\
\cite{kuchinski} and \cite{burgarella} compared rest-frame UV images with visible ones and made 
 pioneering studies where they showed how asymmetry and concentration evolve with the wavelength for several galaxies. Therefore we must pay attention to compare objects that are actually comparable. More specifically,
 we must understand the behavior of the concentration and the asymmetry as a function of the wavelength before comparing results from samples at different $z$. Finally, a physical interpretation of the morphology and its evolution with $\lambda$ is necessary to understand the formation and the evolution of galaxies through high-$z$ observations, and to constrain models.\\
This work relies on the largest good-quality data presently available from UV to visible wavelengths. The paper aims at describing how we can define a spectro-morphology classification from the local sample to $z\sim1$ galaxies.  This work outlines the main parameters and analyses their relative strength and weakness.\\
This paper is organized as follows: the data are presented in Section 2. We discuss the concentration and asymmetry calculations in Section 3. Section 4 presents the results of multi-$\lambda$ measurements. We calibrate these parameters from the nearby sample; and in part 4.1 we give an example of application at high $z$. Finally, we present in Section 5 our conclusions and perspectives.

\section{Data}

\subsection{Sample selection}
A preliminary task is to gather the largest multi-$\lambda$ sample of nearby galaxies from UV to I-band. The sample of galaxies is listed in Table~\ref{tab:set}. The main constraint on building a multi-wavelength sample is the availability of UV data. We use all the UV data in the local
Because of UV observations, the sample is strongly biased against early type galaxies, which are under-represented. We use all the UV data in the local universe available et the time of the work to built a sample weighted toward spiral and irregular galaxies and toward non-active galaxies. There is no other selection. \\
The UV data at 1500 \AA \ and 2500 \AA \ mainly comes from the Ultraviolet Imaging Telescope (UIT, Marcum et al. 2001), with a pixel size of 1.14 arcsec/pixel and a $FWHM$ of about 3.3 arcsec. We have completed the UV sample with data from the FOCA telescope (Milliard et al. 1992), that were observed at 2000 \AA \ with a pixel scale of 3.44 arcsec/pixel and $FWHM=12''$ for M100; a pixel size of 5.2 arcsec/pixel and $FWHM=20''$ for M81.\\

\begin{figure*}[htbp]
\centering
\includegraphics[height=24cm]{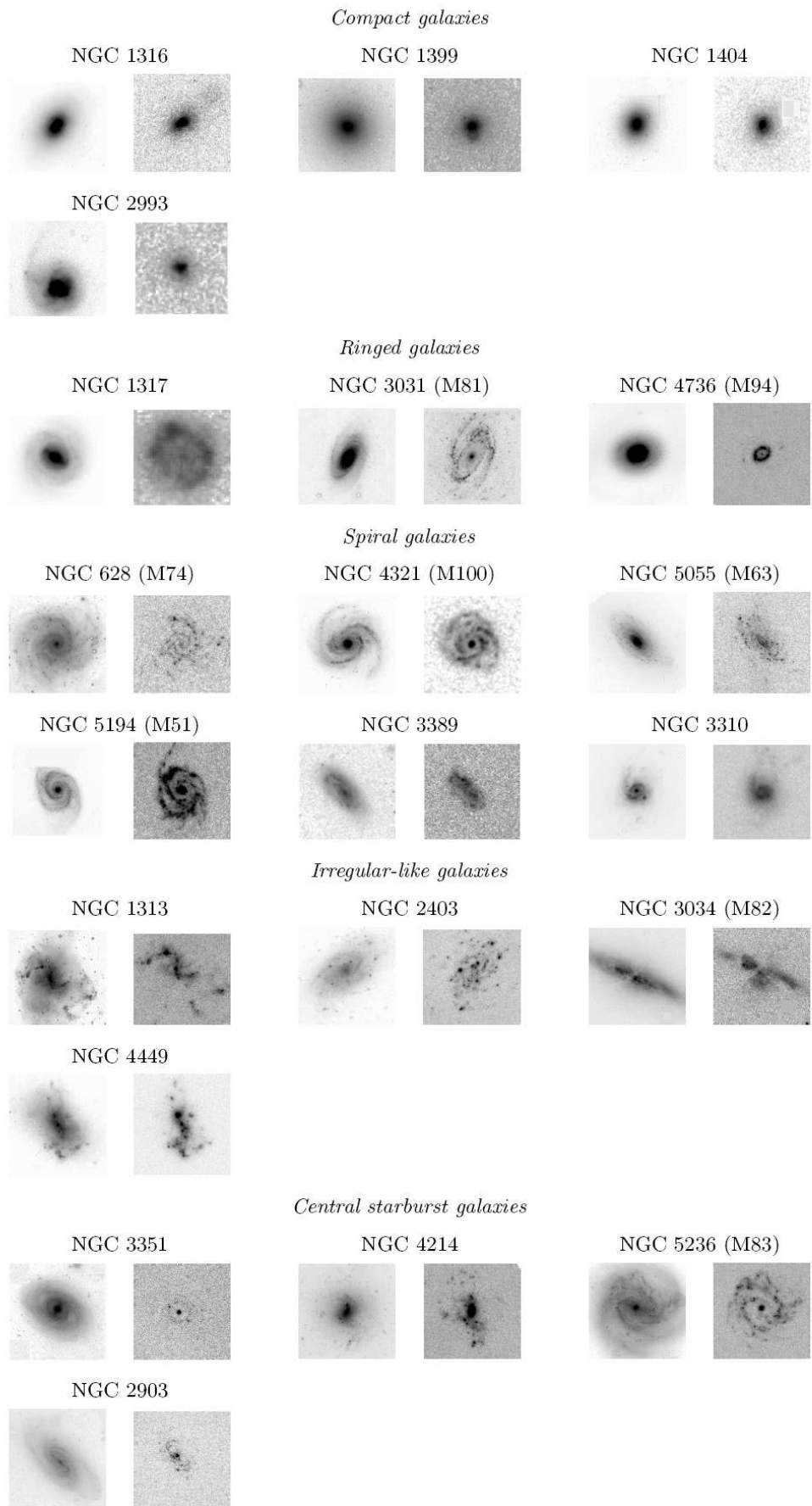}
\caption{Galaxies of our sample classified as a function of their spectro-morphological type. For each galaxy: left: optical image; right: UV image.}
\label{fig: pictures}
\end{figure*}

The pixel scales for optical images range from 0.35 arcsec/pixel to 1.40 arcsec/pixel in visible with a spatial resolution varying from 1.5 arcsec to 3.3 arcsec. Their origin is recapitulated in Table~\ref{tab:set}.
Physical properties of these galaxies are recapitulated in Table~\ref{tab: results} and are illustrated in Figure~\ref{fig: pictures}.

\section{Morphological parameters}

The photometry was performed using the IRAF/ELLIPSE software.
The apertures were centered on the the most intense pixel of the galaxies.
The ellipse parameters for each galaxy were estimated by fitting  the isophote at 3 $\sigma_{noise}$ over the background. 

\subsection{Surface brightness}

Bershady et al. (2000) defined a total photometry aperture using the local surface brightness. More specifically they introduced an undimensional parameter $\eta$, as the ratio between the local surface brightness at the semi-major axis $r$ and the average surface brightness within $r$: $\eta(r) = I(r)/<I(r)>$. They defined the total photometry aperture as twice the semi-major axis where $\eta(r) = 0.2$. The total flux measured at this radius is similar to that estimated when using the curve of growth. Since it is based on a ratio of two fluxes, $\eta$ is not very sensitive to the total luminosity of the galaxy, and allows us to use it on faint-luminosity galaxies. According to \cite{bershady}, for an exponential profile, more than 99\% of the flux is included within $r(\eta=0.2)$, and about 89\% for an $r^{1/4}$-law profile, because of the slower decrease at large radii. For comparison, the ratio between $r(\eta=0.2)$  and the half-light radius is 2.16 and 1.82 respectively for exponential and $r^{1/4}$-law profiles. In the following of the paper, we adopt the method of Bershady et al. to estimate the total photometry aperture of galaxies.

\subsection{Concentration of light}
\label{C}
Kent (1985) defined a parameter of concentration ($C$) based on the ratio of two isophotal radii: the first one contains 80 \% of the total flux of the galaxy, the second one contains 20 \% of that flux. The expression of Kent's parameter is: 
$$C_K = 5 \log(\frac{r_{80 \%}}{r_{20 \%}}).$$
The inner and outer radii are chosen in order to optimize the dynamical range in concentration and to be sensitive enough to the variations of morphological types. Theoretically, a $r^{\frac{1}{4}}$-law profile corresponds to a concentration value $C = 5.2$, and an exponential profile to $C = 2.7$. \cite{bershady} showed that $C$ is very stable against a spatial resolution degradation and thus appears as a robust parameter for the study of high redshift galaxies. We simulated images by degrading the signal-to noise ratio and the spatial resolution: the mean error on $C$ is estimated to be ${\Delta}C\sim0.2$. 
Hereafter we adopt the concentration of Kent as our parameter of concentration.\\

\subsection{Asymmetry}
\label{A}
\subsubsection{Definition}
The asymmetry index was first defined by \cite{schade95}:
$$
A_S = \frac{1}{2} \frac{\sum|I_0-I_{180^o}|}{\sum I_0}.
$$
as a measure of the asymmetry of galaxies by a $180^o$ rotation. 
It consists in computing the pixel by pixel difference between the original image and its 180$^o$ rotation. In the expression of $A_S$, $I_0$ is the intensity for a given pixel and $I_{180^o}$ is the intensity of its symmetric after the rotation. The sky background is subtracted and a careful flat-fielding must be performed before any calculation. The rotation centre is the position that yields a minimum value for the asymmetry.\\ 
\cite{kuchinski} computed the asymmetry only on the pixels above a threshold of light intensity, in order to cancel the effect of pixels that do not contain galaxy light. Indeed, the number of those pixels within the aperture can be important, particularly on UV images of galaxies, that often have a patchy appearance. We used the following definition of asymmetry, inspired from \cite{kuchinski}:
$$
A = \frac{1}{2}\frac{\sum_{I>n\sigma_{sky}}|I-I_0|}{\sum I_0}
$$

where
$n\sigma_{sky}$ is the threshold above the sky background. We chose $n=2$.\\
A correction term must be added to the asymmetry, to remove the background asymmetry. \cite{abraham96} and \cite{conselice} proposed to select a patch of sky with the same size than the aperture to compute its asymmetry. They used the following correction term:
$$ A_{sky} = \frac{1}{2} \frac{\sum|B_0-B_{180^o}|}{\sum I_0} $$
where $B_0$ and $B_{180^o}$ are the intensities of a given pixel from the sky and its rotational symmetric.
However, this method requires to have enough sky on the image around the galaxy; and the time of computation is rather long. We try to avoid these caveats by assuming that the sky noise is purely poissonian and calculated the noise asymmetry in a statistical way:
$$ \sum|B_0-B_{180^o}| = \frac{2}{\sqrt{\pi}}\sigma_{sky}N_{pix}. $$
We tested that the sky noise asymmetry we obtained by the statistical way is very close to those obtained with the method of \cite{abraham96} and \cite{conselice}. Depending on the quality of images, the uncertainty on the sky noise asymmetry can vary from ${\Delta}A_{sky}=\pm0.001$ to $\pm0.01$.\\
Hereafter we adopt the following definition of the asymmetry:
$$
A = \frac{1}{2} [\frac{\sum_{I>n\sigma_{sky}}|I-I_0|}{\sum I_0}-k_{scale}\frac{2}{\sqrt{\pi}}\frac{\sigma_{sky}N_{pix}}{\sum I_0}]
$$
where:
\begin{itemize}
\item $N_{pix}$ is the number of pixels within the aperture;
\item $k_{scale}$ is a scaling factor equal to the ratio of the number of pixels used for computing $A$, to the number of pixels in the patch of sky.
\end{itemize}
We compared them to those obtained by \cite{conselice} when studying images in B-band of a sample of 113 nearby bright galaxies (Frei et al. 1996). 
From this comparison we estimated that the uncertainty on $A$ to be of the order of ${\Delta}A=\pm0.02$.
\subsubsection{Condition of validity of asymmetry}
\begin{figure*}[ht]
\centering
\includegraphics[width=13cm]{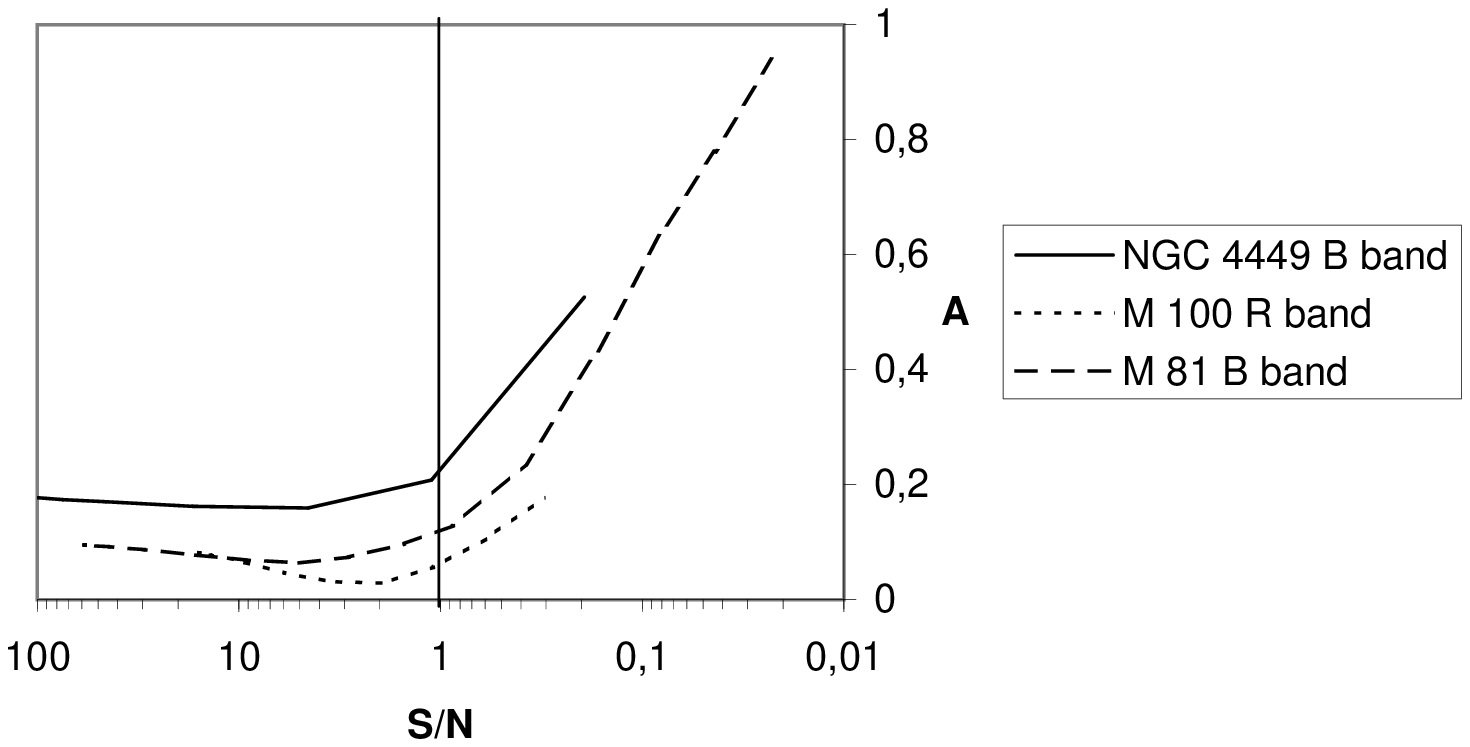}
\caption{Variations of $A$ as a function of the signal-to-noise per pixel within $r(\eta=0.2)$. There is a conservative limit at $S/N_{pix}=1$ for which the method can be safely applied. One pixel covers about 0.2 kpc.}
\label{fig: A_vs_sn}
\end{figure*}

The asymmetry index mainly depends on two factors: the spatial resolution and the signal-to-noise ratio per pixel $S/N_{pix}$. 
\cite{conselice} defined a minimum spatial resolution, that is linked to the size of star-forming regions. They introduced $\epsilon=\frac{\theta_{0.5kpc}}{\theta_{res}}$ the ratio of the angular size of 0.5~$h^{-1}_{75}$~kpc in the galaxy to the spatial resolution. They showed by degrading the pixel scale that below $\epsilon\sim1$, asymmetry values fall, galaxies becoming too symmetric: morphological details such as star-forming regions are not resolved anymore. 
Nevertheless \cite{conselice} estimated that for large galaxies a resolution of 1 kpc can be acceptable. Indeed, some acceptable values of $A$ found for images at $\epsilon<1$ (\textit{e.g.} M 100 in UV taken by FOCA with $\epsilon\simeq0.6$) suggest that this constraint on spatial resolution is not absolute. 
We verified that for all the images except the FOCA image of M 100 the condition defined by \cite{conselice} is satisfied.\\
Once the physical constraint on the spatial resolution secured, the other decisive factor, of statistical nature, is $S/N_{pix}$. 
In order to test its effect we degraded $S/N_{pix}$ for several galaxies of various types and various wavelengths, previously resized at the same physical scale: $\epsilon\sim2.7$ \textit{i.e.} one pixel covers about 0.2 kpc in the galaxy. The asymmetry shows low changes as long as $S/N_{pix}\ge1$ (Figure~\ref{fig: A_vs_sn}). When $S/N_{pix}\le1$, large diffuse regions of the galaxy become invisible. Only the brightest structures are seen; because of the selection of the brightest pixels, few are used in the computation of $A$. As a result, $A$ increases. Therefore we must have $S/N\ge1$ to compute safely $A$. This statistical condition is valid whatever the pixel scale if the condition on the spatial resolution $\epsilon\ge1$ is verified. All the data compiled (except NGC 3351 at 1500\AA \ which was not used for asymmetry computation) verified the physical constraint on the spatial resolution and the statistical condition on $S/N_{pix}$, \textit{i.e.} $\epsilon\ge1$ and $S/N_{pix}\ge1$.


\section{Results}
\subsection{Bandshifting effects}
\label{agn}

Measuring the variation of asymmetry and concentration indices as a function of $\lambda$ is equivalent to observe the distribution of different types of stellar populations emitting at different wavelengths within a galaxy. Thus, we might be able to distinguish different types of galaxies. Star-forming regions and young blue stars are prominent in rest-frame (RF) UV whereas older stellar population are observed in RF visible. 
Consequently, the morphology of galaxies at these two wavelengths could extremely change in the case of two-component stellar population galaxies, like spiral galaxies. 
Concentration values in UV are expected to be lower than in optical, for the same galaxy (Kuchinski at al. 2000). 
Indeed, some early spiral galaxies NGC 1317, M81, and M94 
present a ring of young stars in UV (Reichen et al. 1994, Waller et al. 2001), with a faint or nonexistent central bulge that trends to decrease the concentration value. In R-band, they are very concentrated because of their prominent central bulge of red old stars. Thus, that significant change in concentration is due to the coexistence of two stellar populations that are differently distributed. The change in concentration is less dramatic toward late type galaxies: their star-forming regions are more or less uniformly distributed in all the galaxy, and their surface brightness profile is well fitted by an exponential profile. Irregular galaxies like NGC 4449 and M82 keep a constant concentration in UV and visible, that is the expression of only one young stellar component in these galaxies.\\
The clumpy appearance of late-type spirals in UV, due to the presence of star-forming regions, gives higher values of $A$ in UV than in optical wavelengths, where the disk appears smoother because of a more uniform repartition of old stars.

\subsection{Spectro-morphological types}

\subsubsection{Qualitative definitions}
\begin{figure*}[htbp]
\centering
\resizebox{\hsize}{!}{\includegraphics{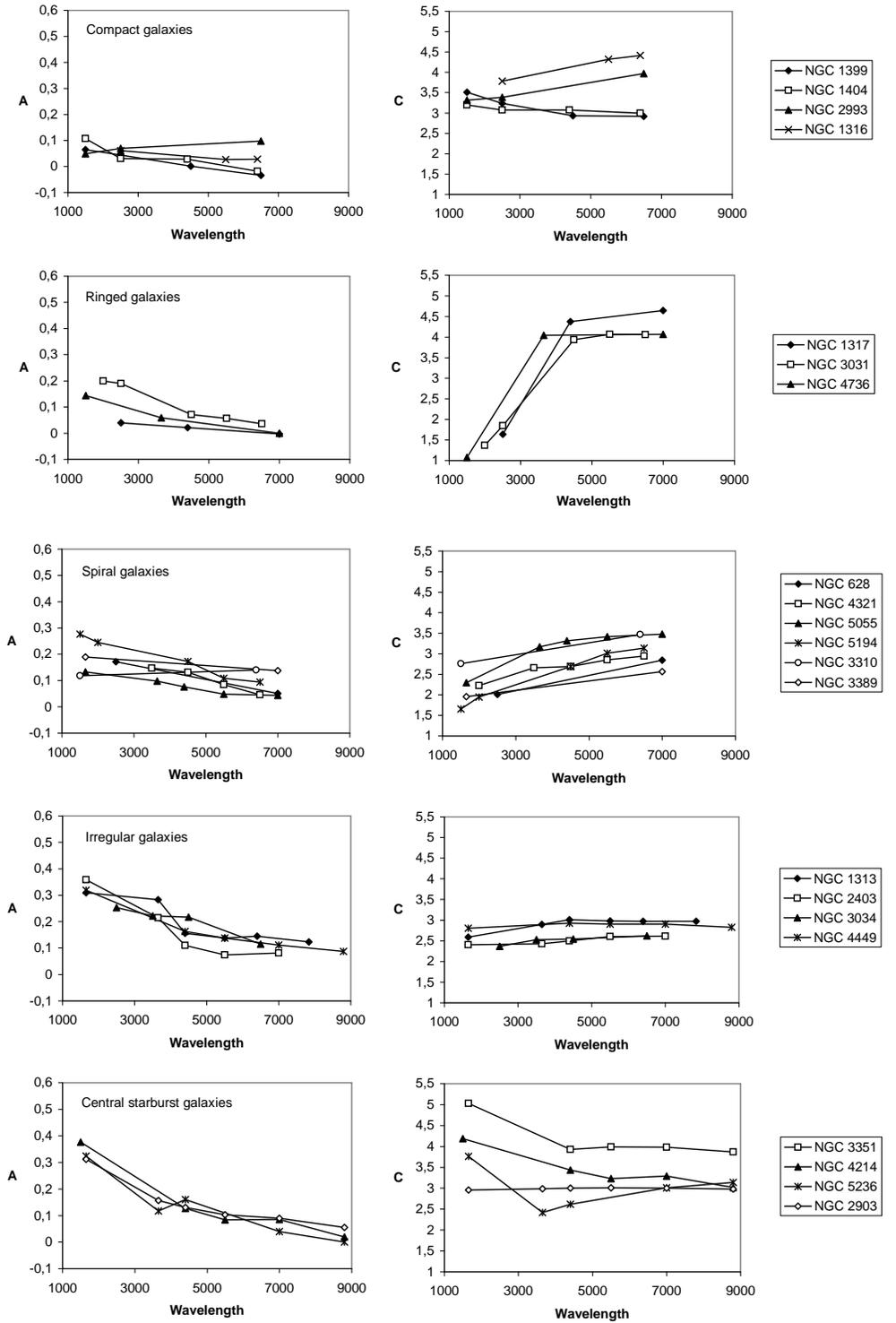}}
\caption{Spectra of concentration and asymmetry versus $\lambda$ in \AA.}
\label{graph: profil}
\end{figure*}

Previous works (Abraham et al. 1996, Bershady et al. 2000) showed that it is possible to discriminate three main morphological types of galaxies (E/SO, spiral and irregular/merger) from an asymmetry-concentration diagram built in R or B-band. This is no more possible in UV, as shown by \cite{burgarella} and \cite{kuchinski}: at these wavelengths, spiral galaxies appears with a later type and we are not able to distinguish them from irregular galaxies. Therefore one-band classifications only account for part of the available information and would lead to misinterpretations at high redshift ($z\ge1$). A multi-wavelength study would allow us to better define morphological types which would account for some elements of physics of galaxies, such as repartition and proportion of stellar populations and star-forming activity.\\
In Figure~\ref{graph: profil} are reported the asymmetry and concentration parameters of the galaxies in our multi-band sample and both parameters as a function of wavelength. Five generic behaviors were observed, that we will call hereafter spectro-morphological types:
\paragraph{Compact galaxies:} These galaxies are characterized by a low asymmetry value whatever the wavelength; and globally high values of concentration, with low changes with the wavelengths. This spectro-morphological type includes E/SO galaxies, and early spiral galaxies dominated by their central bulge at all wavelengths. The low changes in asymmetry and concentration is the result of the presence of only one stellar component in these galaxies. 
\paragraph{Ringed galaxies:} These galaxies feature in UV a faint or inexistent central bulge surrounded by a bright ring of stellar formation activity. This ring is often induced by orbital resonance with the central bulge or a stellar bar, and likely takes place at the ILR of these galaxies (\textit{e.g.} NGC 4736). Rings might also be due to a past merger activity, or gravitational interaction. Ringed galaxies show a prominent central bulge of old stars at optical wavelengths. This dramatic difference in the repartition of young and old stars yields high changes in the concentration with the wavelength. The change in asymmetry is not very significant because of the symmetry of the ring structure.
\paragraph{Spiral galaxies:} Their changes in concentration and asymmetry is the result of the existence of a composite stellar population. However their appearance changes less dramatically than ringed galaxies with the wavelength; they keep their spiral-like structure whatever the wavelength. Star-forming regions are located on the spiral arms and give a patchy appearance in UV whereas old stars form a prominent bulge and a uniform disk.
\paragraph{Irregular galaxies:} The main feature of these galaxies is the nil change in concentration with the wavelength, and the concentration value expected is close to 2.7 \textit{i.e.} the theoretical value for an exponential surface brightness profile. These galaxies show a prominent bulge neither in UV nor in visible; their main stellar component is young. Their patchy appearance in UV gives high asymmetry values at this wavelength, and decrease toward longer wavelengths because of their more diffuse structure in optical bands. Late spiral and irregular galaxies, with low surface brightness or flocculent disk are not distinguished and both constitute the irregular spectro-morphological type.
\paragraph{Central starburst galaxies:} Galaxies with a central starburst are easily recognizable from their particularly high concentration value in UV, and a positive change in concentration from UV to R-band. As suggered by \cite{colina} and \cite{scoville}, there is likely a strong link between nuclear starburst activity and AGN. We expect to find in this class some active galaxies.\\

\subsubsection{Spectro-morphological boxes}
\begin{figure*}[htbp]
\centering
\subfigure[\label{graph: diagAC:A}]{\includegraphics[height=5cm]{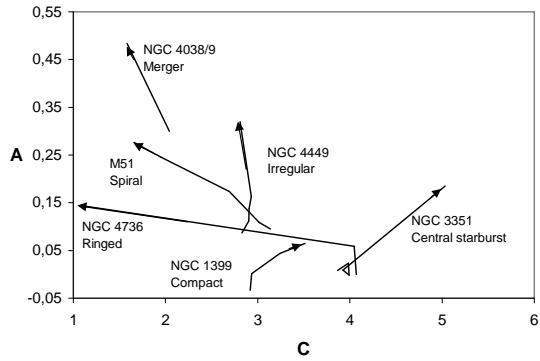}}
\vspace{0.5cm}
\subfigure[\label{graph: diagAC:B}]{\includegraphics[height=5cm]{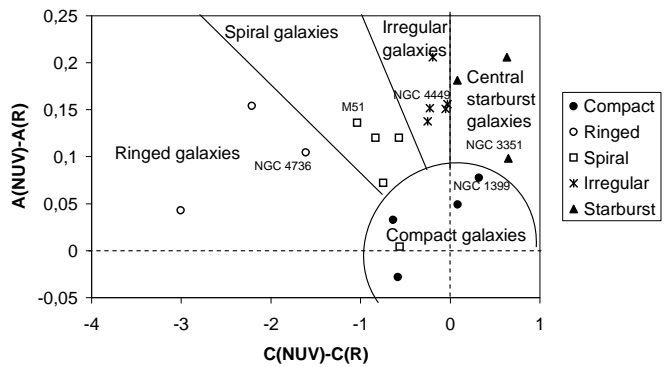}}
\caption{\textit{(a)}: shift of some representative galaxies of each spectro-morphological type from R-band to UV. The arrows point toward UV. The location of NGC 4038/9 on the diagram is well distinct of the location of single systems. \textit{(b)}: change in $C$ versus change in $A$. It is possible to discriminate five spectro-morphological area: compact, ringed, spiral, irregular and central starburst.}
\label{graph: diagAC}
\end{figure*}

Quantitatively, we are able to define spectro-morphological boxes, based on $A(R$), $C(R)$, and the changes in $A$ and $C$ with $\lambda$, hereafter ${\Delta}A$ and ${\Delta}C$ (Table~\ref{tab: mean values}). The elliptical galaxies NGC 1399 and NGC 1404 which are not very concentrated may well not be very representative of this morphological class. A larger sample of early galaxies is necessary to improve the analysis, but we expect the general trends to be true.
Each spectro-morphological type is characterized by a distinct shift in the asymmetry-concentration diagram when studied in visible and UV. We illustrate this effect in Figure~\ref{graph: diagAC:A} by plotting one typical case for each spectro-morphological type.
For the sake of comparison, we also reported the locus of the merger NGC 4038/9.
Its changes in concentration and asymmetry with the wavelength are the result of the existence of a composite stellar population. The high values of the asymmetry whatever the wavelength allow us to discriminate it from single systems. Larger samples will be necessary to generalize this behavior for mergers.
In Figure~\ref{graph: diagAC:B} is also plotted ${\Delta}A$ versus ${\Delta}C$  for our sample. This spectro-morphological diagram is a projection of two of the four parameters $A(R$), $C(R)$, ${\Delta}A$ and ${\Delta}C$ that define a spectro-morphological type.
We can empirically define five areas in this diagram, corresponding to the five spectro-morphological types defined above. In summary, when ${\Delta}C$, galaxies are classified as central starburst. When $\sqrt{{\Delta}A^2+{\Delta}C^2}<1$, galaxies are classified as compact galaxies. UV-ringed galaxies, spiral galaxies and irregular galaxies areas are delimited by the following equations : ${\Delta}A=-0.31{\Delta}C$ and ${\Delta}A=-0.088{\Delta}C$, whith ${\Delta}C<0$.

\subsection{First application to high-$z$ galaxies and perspectives}

\begin{figure*}[htbp]
\centering
\resizebox{\hsize}{!}{\includegraphics{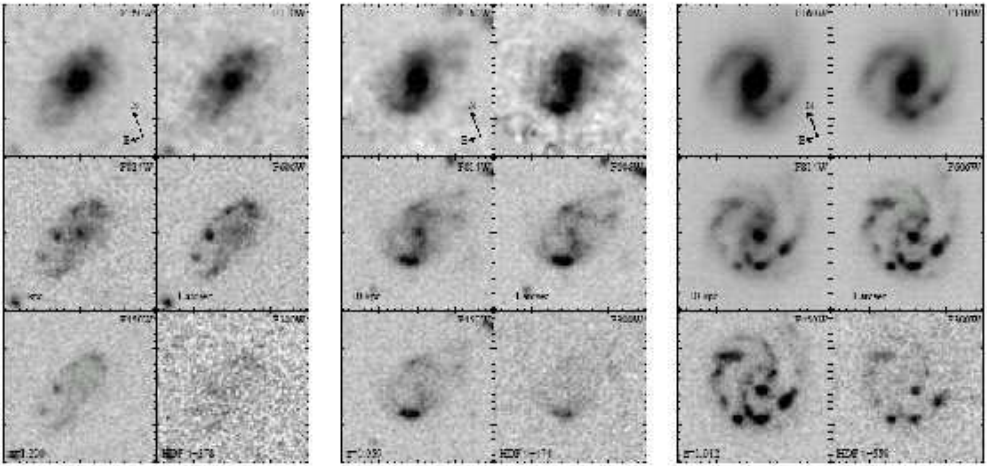}}
\caption{Multi-wavelength images of the three galaxies HDF 4-378, 4-474 and 4-550. The F160W images (RF I-band) let appear a prominent bulge and the spiral structure of disk, at least for HDF 4-474 and 4-550. At shorter wavelengths appear star-forming regions. The F300W images were removed from the study because of too low signal-to-noise ratio.}
\label{images hdf}
\end{figure*}

The application of spectro-morphology to high$-z$ galaxies can make use of deep surveys which provide multi-$\lambda$ data. 
As an example, we have computed the morpho-spectra of three obviously-spiral galaxies from the HDF (Bunker et al. 2000; Figure~\ref{images hdf}): HDF 4-378, 4-474 and 4-550, respectively situated at $z=1.2$, $z=1.059$ and $z=1.012$ and observed by HST at six wavelengths, from the RF FUV to I-band. F300W, F450W, F606W and F814W-filter images were provided by the HST Wide-Field Camera 2 with a pixel size of 0.045 arcsec, \textit{i.e.} $\epsilon\simeq1.3$.
F110W and F160W-filter images provided by NICMOS (NIC3) have a pixel size of 0.096 arcsec, \textit{i.e.} $\epsilon\simeq0.7$. 
The values of $\epsilon$ for the NIC3 images are only slightly lower than the threshold recommended by \cite{conselice}. As noticed in Section \ref{A} this threshold is not absolute, and the values of $A$ found with the NIC3 images are considered as acceptable.
We used the F450W, F606W, F814W, F110W and F160W images which correspond respectively to the RF near-UV, U, B, V and I-bands. The signal-to-noise ratio per pixel of each image is larger than 1.
The changes in $C$ and $A$ (Table~\ref{hdftab}) found for HDF 4-378 and 4-474 are typically those found for spiral  galaxies, whereas both galaxies were classified as irregular by \cite{fernandez-soto} from their spectral energy distributions (SED). 
For HDF 4-550, the change in asymmetry is quite high, due to the presence of particularly bright star-forming regions on the arms of that galaxy. The higher the changes in $C$ and in $A$ with the wavelength are, the more dramatic the contrast in the repartition of young and old stellar populations is.
This example of application to high redshift galaxies is successful. However, we must understand the influence of active-nuclear and interacting objects, that we expect to be more important at larger distances. AGN galaxies would likely be classified as central starburst galaxies in our classification, due to the central activity that often occurs in the central bulge (\cite{maiolino}). Moreover we have to focus on the application of spectro-morphology to merger systems, expected to be more numerous at high redshift. One of the main results of this method would be to avoid misclassifying spiral galaxies into irregulars. Indeed, high-$z$ observations often seem to suggest that there is an increase of irregular galaxies with the redshift and a low contribution of spiral galaxies which are strong constraints to models of galaxy formation. So, it is important to discriminate spiral and irregular galaxies. A multi-$\lambda$ view will make it possible to use all the information contained in the images and minimize misclassifications of spirals into irregulars.\\
A more detailed study of high-redshift galaxies is devoted to a next paper. Interacting systems and AGN galaxies, inducing star formation activity at the centre of these galaxies, are expected to be more common than in the local universe. The spectro-morphology relies on stellar populations. By studying the repartition of stellar populations we could bring observational constraints on the scenarios of evolution of galaxies.

\section{Conclusion}

A multi-wavelength study of a sample of nearby objects  allowed us to elaborate what we call spectro-morphology, \textit{i.e.} the study of the behavior of asymmetry and concentration with the wavelength. We saw that two-stellar component galaxies (spiral galaxies of the Hubble classification) present the highest variation of both indices, whereas the variation of one-stellar component galaxies (ellipticals/lenticulars and irregulars) are less wavelength-dependent. 
More precisely these changes in concentration and in asymmetry with the wavelength between UV and R-band led us to define five spectro-morphological types, characterized by their repartition of stellar populations.
Compact galaxies are concentrated and symmetric at all wavelengths. This constancy in $C$ and $A$ is the result of the presence of only one stellar component. Ringed galaxies are often early spiral galaxies with a prominent bulge in visible which exhibits a ring of star-forming activity in UV: the locus of young and old stellar populations are very different, which induce high changes in $C$ and $A$ from UV to optical bands. Spiral galaxies show a prominent bulge in visible and star-forming regions on the arms of the spiral structure in UV. The resulting changes in $A$ and $C$ with the wavelength are less dramatic than for ringed galaxies, but characterize the existence of a composite stellar population. Irregular galaxies have a concentration parameter close to 2.7 at all wavelengths, that is the value expected for a galactic disk, and are rather asymmetric. This class contains galaxies with a faint or inexistent central bulge, which are classified in the Hubble classification as late spiral or irregular. Their stellar population is dominated by young stars. Central starburst galaxies possess a very bright nucleus in UV that induces a high concentration. This starburst activity is often induced by gravitational interaction or the existence of an AGN.
The correlation between $C(NUV)-C(R)$ and $A(NUV)-A(R)$ allowed us to identify these five spectro-morphological types. 

To test the validity of our method at high redshift, we successfully applied it to three distant objects in the HDF at $z\sim1$. The resulting morphology is consistent with an eye-ball morphology and these galaxies are classified as spiral galaxies. 


\begin{acknowledgements}
We thank Andrew Bunker for his collaboration, Rodger I. Thompson for giving authorization to use the HDF multi-wavelength images. We are also grateful to the anonymous referee who largely contributes to improve this work. The support given by ASTROVIRTEL, a Project funded by the European Commission under FP5 Contract No HPRI-CT-1999-00081 is acknowledged. 
\end{acknowledgements}

\newpage
\vspace*{7cm}

\newpage


\begin{landscape}
\begin{table*}[ht]
\centering
\begin{tabular}{lcccccccccc}
\hline
Galaxy & 1500 \AA & 2000 \AA & 2500 \AA & U & B & V & R & I & UV data & Optical data\\
\hline
\hline
NGC 628 & & & X & & & & X & & UIT (1)& Palomar P60, 0.378 arcsec/pixel (6)\\ 
NGC 1313 & X & & & X & X & X & X & X & UIT (1) & Danish 1.54m, La Silla, 0.4 arcsec/pixel (2)\\
NGC 1316 & & & X & & & X & X & & UIT (1) & Cerro Tololo Inter-American Observatory (CTIO) (1)\\ 
NGC 1317 & & & X & & X & & X & & UIT (1) & CTIO (1)\\ 
NGC 1399 & X & & X & & X & & X & & UIT (1) & Mount Laguna Observatory (MLO) (1) \\
NGC 1404 & X & & X & & X & & X & & UIT (1) & ESO archive (7)\\
NGC 2403 & X & & & X & X & X & X & & UIT (1) & P60 (3)\\
NGC 2903 & X & & & X & X & X & X  & & UIT (1) & P60 (3)\\ 
NGC 2993 & X & & X & & & & X & & UIT (1) & MLO (1)\\ 
NGC 3031 & & X & X & & X & X & X & & FOCA (8) \& UIT (1) & Observatoire de Haute-Provence (OHP), 0.69 arcsec/pixel (4)\\ 
NGC 3034 & & & X & X & X & & X & & UIT (1) & OHP (4)\\ 
NGC 3310 & X & & & & & & X & & UIT (1) & Jacobus Kapteyn telescope (JKT), la Palma (5)\\ 
NGC 3351 & X & & & & X & X & X & X& UIT (1) & P60 (3)\\
NGC 3389 & X & & & & & X & & & UIT (1) & P60 (3)\\ 
NGC 4038/39 & X & & & & & & X & & UIT (1) & ESO archive (7)\\
NGC 4214 & X & & & & X & X & X & X & UIT (1) & P60 (3)\\ 
NGC 4321 & & X & & X & X & X & X & & FOCA (8) & OHP (4)\\ 
NGC 4449 & X & & & & X & X & X & X & UIT (1) & P60 (3) \\
NGC 4736 & X & & & X & & & X & & UIT (1) & P60 (3)\\ 
NGC 5055 & X & & & X & X & X & X & & UIT (1) & P60 (3)\\
NGC 5194 & X & X & & & X & X & X & & UIT (1) \& FOCA (8) & OHP (4)\\
NGC 5236 & X & & & X & X & & X & X & UIT (1) & CTIO (3)\\

\hline
\end{tabular}
\caption{Data used for the calibration of the method. (1): \cite{marcum}.
(2): \cite{larsen}.
(3): \cite{kuchinski}.
(4): Boselli et al. (private communication).
(5): \cite{james}.
(6): Palomar P60; 0.378 arcsec/pixel; Madore et al. (private communication).
(7): Schmidt telescope 1m; 2.5 arcsec/pixel; \cite{lauberts}.
(8): \cite{milliard}.
}
\label{tab:set}
\end{table*}
\end{landscape}

\begin{landscape}
\begin{table*}[htbp]
\begin{tabular}{lccccccccc}
\hline
Galaxy & Type t $^1$ & RC3 Type $^2$& Physical peculiarities & $A(R)$ & $C(R)$ & $A(UV)-A(R)$$ ^3$ & $C(UV)-C(R) $$^3$ & $r(\eta=0.2)_R $$^4$ &  $M_B $$^1$  \\
\hline
\hline
NGC 628 / M 74 \dotfill & 5.2 & SA(s)c & \dots & 0.051 & 2.84 & 0.12 & -0.83 & 212  & -20.29\\ 
NGC 1313 \dotfill & 6.9 & SB(s)d &  \dots  & 0.145 & 2.97 & 0,152 &-0,23& 232  & -19.12 \\      
NGC 1316 \dotfill & -1.9 & SAB(s)0 & pec & 0.028 & 4.42 & 0,033 & -0,63 & 268  & -22.07  \\ 
NGC 1317 \dotfill & 0.7 & SAB(r)a & ring UV & -0.003 & 4.64 & 0,043 & -3,01 & 66  & -20.16 \\ 
NGC 1399 \dotfill & -4.9 & E1 & pec & -0.034 & 2.92 & 0,077     & 0,32 & 168  & -20.95  \\
NGC 1404 \dotfill & -5 & E1 &  \dots & -0.018 & 2.99 & 0,049 & 0,08 & 132  & -21.13 \\
NGC 2403 \dotfill & 5.8 & SAB(s)cd & \dots  & 0.081& 2.61& 0,206 & -0,20 & 316  & -19.52 \\
NGC 2903 \dotfill & 4 & SAB(rs)bc &  sbrst & 0.090& 3.00& 0,146 & -0,03 & 230  & -20.90 \\

NGC 2993 \dotfill & 1.1 & Sa & pec, sbrst & 0.097 & 3.97 & -0,028 & -0,58 & 25  & -19.72 \\

NGC 3031 / M 81 \dotfill & 2.2 & SA(s)ab & ring UV, LINER, Sy 1.8 & 0.036 & 4.06 & 0,154 & -2.21 & 457  & -21.54 \\
NGC 3034 / M 82 \dotfill & 10 & I0 &  \dots & 0.115 & 2.62 & 0.137 & -0.25 & 173  & -18.52\\ 
NGC 3310 \dotfill & 4 & SAB(r)bc & pec, sbrst & 0.140 & 3.47 & -0,016   & -0,56 & 30  & -20.25\\

NGC 3351 / M 95 \dotfill & 2.9 & SB(r)b & ring UV, sbrst & 0.023 & 3.98 & \dots & 0,65 & 166  & -20.30 \\

NGC 3389 \dotfill & 2.3 & SA(s)c &  \dots & 0.138 & 2.56 & 0,041 & -0,49 & 76 & -19.79 \\
NGC 4038/9 \dotfill & 99 & merger & \dots & 0.299 & 2.04 & 0.145 & -0.36 & 165  & -21.34/-21.30 \\
NGC 4214 \dotfill & 10 & IAB(s)m &  \dots & 0.085 & 3.30 & 0,206 & 0,63 & 153  & -17.11 \\

NGC 4321 / M 100 \dotfill & 4.3 & SAB(s)bc &  \dots & 0.047 & 2.94 & \dots & -0,57 & 190  & -22.12 \\

NGC 4449 \dotfill & 10 & IBm &  \dots & 0.112 & 2.90 & 0,151 & -0,05 & 127  & -17.80 \\

NGC 4736 / M 94 \dotfill & 2.1 & (R)SA(r)ab & LINER, Sy 2 & 0.000 & 4.07 & 0,105 & -1,61 & 49  & -21.06 \\

NGC 5055 / M 63 \dotfill & 4 & SA(rs)bc & \dots & 0.043 & 3.48 & 0,072 & -0,75 & 293  & -21.22 \\

NGC 5194 / M 51 \dotfill & 4.2 & SA(s)bc & pec, Sy 2.5 & 0.095 & 3.14 & 0,136 & -1,06 & 221  & -19.70 \\

NGC 5236 / M 83 \dotfill & 5.1 & SAB(s)c & sbrst & 0.040 & 3.01 & 0,181 & 0,08 & 353  & -21.12\\
 
\hline
\end{tabular}
\caption{Concentration and asymmetry parameters and their variations with the wavelength. $^1$~see \textit{http://leda.univ-lyon1.fr}.
$^2$~Morphological type from the RC3.
$^3$~Computed by interpolation when NUV data is not available, because of the significant dynamics of $A$ and $C$ between 1500~\AA and 3500~\AA. 
$^4$~Value in R-band, in arcsec.} 
\label{tab: results}
\end{table*}
\end{landscape}

\newpage
\vspace*{2cm}

\begin{table*}[ht]
\centering
\begin{tabular}{lcccc}
\hline
Morphological Type & $A(R)$ & $C(R)$ & $A(NUV)-A(R)$$^1$ & $C(NUV)-C(R)$$^1$ \\
\hline
\hline
Compact galaxies \dotfill & $0.018\pm0.059$  & $3.58\pm0.74$  & $0.033\pm0.045$ & $-0.20\pm0.48$\\
Ringed galaxies \dotfill & $0.011\pm0.022$ & $4.26\pm0.33$ & $0.101\pm0.056$ & $-2.28\pm0.70$ \\
Spiral galaxies  \dotfill& $0.086\pm0.046$ & $3.07\pm0.36$ & $0.090\pm0.054$ & $-0.75\pm0.20$ \\
Irregular galaxies \dotfill & $0.113\pm0.026$ & $2.78\pm0.19$ & $0.161\pm0.030$ & $-0.18\pm0.09$ \\
Central starburst galaxies \dotfill & $0.059\pm0.033$ & $3.32\pm0.46$ & $0.160\pm0.046$ & $+0.33\pm0.36$ \\
\hline
NGC 4038/9 & 0.299      & 2.04 & 0.145 & -0.36\\
\hline
\end{tabular}
\caption{Means values of morphological parameters. $^1$~computed at 2500\AA\ and 6500\AA\ rest-frame and by interpolation if these wavelengths are not available. NGC 4038/9 is given as an example of merging system.}
\label{tab: mean values}
\end{table*}
\vspace*{4cm}
\begin{table*}[ht]
\centering
\begin{tabular}{p{2cm}cccccc}
\hline
Object & Spectral type$^1$ & $A(R)$$^2$ & $C(R)$$^2$ & $A(NUV)-A(R)$$^{2,3}$ & $C(NUV)-C(R)$$^{2,3}$ \\
\hline
\hline
HDF 4-378 \dotfill & Irregular & 0,066 & 2,34   & 0,141 & -0,63 \\
HDF 4-474 \dotfill & Irregular & 0,056  & 2,22 & 0,177  & -0,82  \\
HDF 4-550 \dotfill & Scd & 0,064        & 2,50 & 0,332 & -0,89  \\
\hline
\end{tabular}
\caption[ ]{Morphological parameters computed for three high-$z$ galaxies. 
$^1$~is taken from \cite{fernandez-soto}. $^2$~rest-frame wavelength. $^3$~computed at 2500\AA\ and 6500\AA\ rest-frame and by interpolation if these wavelengths are not available.} 
\label{hdftab}
\end{table*}

\begin{thebibliography}{}
\bibitem[Abraham et al. (1994)]{abraham94} 
Abraham, R.~G., Valdes, F., Yee, H.~K.~C., \& van den Bergh, S.\ 1994, 
\apj, 432, 75 
\bibitem[Abraham et al. (1996)]{abraham96} Abraham, R.~G., van den 
Bergh, S., Glazebrook, K., Ellis, R.~S., Santiago, B.~X., Surma, P., \& 
Griffiths, R.~E.\ 1996, \apjs, 107, 1
\bibitem[Baugh, Cole, \& Frenk 1996]{baugh} Baugh, C.~M., 
Cole, S., \& Frenk, C.~S.\ 1996, \mnras, 283, 1361 
\bibitem[Bershady et al. (2000)]{bershady} 
Bershady, M.~A., Jangren, A., \& Conselice, C.~J.\ 2000, \aj, 119, 2645.
\bibitem[Bresolin \& Kennicutt (2002)]{bresolin} Bresolin, F.~\& 
Kennicutt, R.~C.\ 2002, \apj, 572, 838 
\bibitem[Brinchmann et al. 1998]{brinchmann98} Brinchmann, J.~et 
al.\ 1998, \apj, 499, 112 
\bibitem[Bunker et al. (2000)]{bunker} Bunker, A., Spinrad, H., Stern, D., ~et al.\ 2000, astro-ph/0004348
\bibitem[Burgarella et al. (2001)]{burgarella} Burgarella, D., 
Buat, V., Donas, J., Milliard, B., \& Chapelon, S.\ 2001, \aap, 369, 421 
\bibitem[Buta 1995]{buta} Buta, R.\ 1995, \apjs, 96, 39 
\bibitem[Castellani \& Tornambe 1991]{castellani} Castellani, 
M.~\& Tornambe, A.\ 1991, \apj, 381, 393 
\bibitem[Colina et al. 1997]{colina} Colina, L., Garcia 
Vargas, M.~L., Mas-Hesse, J.~M., Alberdi, A., \& Krabbe, A.\ 1997, \apjl, 
484, L41 
\bibitem[Conselice et al. (1998)]{conselice98} Conselice, C., Bershady, M.~A., Dickinson, M., ~et al.\ 
1998, Bulletin of the American Astronomical Society, 30, 1368 
\bibitem[Conselice et al. (2000) ]{conselice} 
Conselice, C.~J., Bershady, M.~A., \& Jangren, A.\ 2000, \apj, 529, 886
\bibitem[de Vaucouleurs et al. 1991]{RC3} de Vaucouleurs, 
G., de Vaucouleurs, A., Corwin, H.~G., Buta, R.~J., Paturel, G., \& Fouque, 
P.\ 1991, Volume 1-3, XII, 2069 pp.~7 figs..~ Springer-Verlag Berlin 
Heidelberg New York  
\bibitem[Dirsch et al. (2003)]{dirsch} Dirsch, B., Richtler, 
T., Geisler, D., Forte, J.~C., Bassino, L.~P., \& Gieren, W.~P.\ 2003, \aj, 
125, 1908 
\bibitem[Drozdovsky et al.  2002]{drozdovsky} Drozdovsky, I.~O., 
Schulte-Ladbeck, R.~E., Hopp, U., Greggio, L., \& Crone, M.~M.\ 2002, \aj, 
124, 811 
\bibitem[Elmegreen, Chromey, \& Warren (1998)]{elmegreen} 
Elmegreen, D.~M., Chromey, F.~R., \& Warren, A.~R.\ 1998, \aj, 116, 2834 


\bibitem[Fern{\' a}ndez-Soto et al. (1999)]{fernandez-soto} Fern{\' a}ndez-Soto, A., Lanzetta, K.~M., \& Yahil, A.\ 1999, \apj, 513, 34 
J
\bibitem[Frei et al. (1996)]{frei} 
Frei, Z., Guhathakurta, P., Gunn, J.~E., \& Tyson, J.~A.\ 1996, \aj, 111, 
174. 
\bibitem[Giavalisco et al. 1996]{giavalisco} Giavalisco, M., 
Livio, M., Bohlin, R.~C., Macchetto, F.~D., \& Stecher, T.~P.\ 1996, \aj, 
112, 369 
\bibitem[Greve et al. 2002]{greve} 
Greve, A., Wills, K.~A., Neininger, N., \& Pedlar, A.\ 2002, \aap, 383, 56 
\bibitem[Harris et al.(2001)]{harris} Harris, J., Calzetti, 
D., Gallagher, J.~S., Conselice, C.~J., \& Smith, D.~A.\ 2001, \aj, 122, 
3046 
\bibitem[Hill et al. 1998]{hill} Hill, R.~S.~et al.\ 1998, 
\apj, 507, 179 
\bibitem[Horellou et al. 2001]{horellou} Horellou, C., Black, 
J.~H., van Gorkom, J.~H., Combes, F., van der Hulst, J.~M., \& 
Charmandaris, V.\ 2001, \aap, 376, 837 
\bibitem[Kent (1985)]{kent} Kent, S.~M.\ 1985, \apjs, 59, 115
\bibitem[Knapen et al. 1995]{knapen} Knapen, J.~H., Beckman, 
J.~E., Heller, C.~H., Shlosman, I., \& de Jong, R.~S.\ 1995, \apj, 454, 623 
\bibitem[Kuchinski et al. (2000)]{kuchinski} Kuchinski, L.~E., Madore, B.~F., Freedman, W.~L., Trewhella, M.\ 2000, \apjs, 131, 441 
\bibitem[James et al. (2003)]{james} James, P.~A., Shane, N.~S., Beckman, J.~E., et al., astro-ph/0311030
\bibitem[Lamers et al. 2002]{lamers} Lamers, H.~J.~G.~L.~M., 
Panagia, N., Scuderi, S., Romaniello, M., Spaans, M., de Wit, W.~J., \& 
Kirshner, R.\ 2002, \apj, 566, 818 
\bibitem[Larsen \& Richtler(1999)]{larsen} Larsen, S.~S.~\& 
Richtler, T.\ 1999, A\&A, 345, 59 
\bibitem[Lauberts \& Valentijn (1989)]{lauberts} Lauberts, A.~\& 
Valentijn, E.~A.\ 1989, Garching: European Southern Observatory, |c1989,  
\bibitem[Maiolino et al. 1999]{maiolino} Maiolino, R., 
Alonso-Herrero, A., Anders, S., Quillen, A., Rieke, G.~H., \& 
Tacconi-Garman, L.~E.\ 1999, Advances in Space Research, 23, 875 
\bibitem[Marcum et al. (2001)]{marcum} Marcum, P.~M., O'Connell, R.~W., Fanelli, M.~N., ~et al.\ 
2001, \apjs, 132, 129. 
\bibitem[Milliard et al.(1992)]{milliard} Milliard, B., Donas, 
J., Laget, M., Armand, C., \& Vuillemin, A.\ 1992, \aap, 257, 24 
\bibitem[Morgan(1958)]{morgan58} Morgan, W.~W.\ 1958, \pasp, 70, 
364 
\bibitem[Morgan(1959)]{morgan59} Morgan, W.~W.\ 1959, \pasp, 71, 
92 
\bibitem[Mu{\~ n}oz-Tu{\~ n}{\' o}n, Caon, \& 
Aguerri(2004)]{munoz-tunon} Mu{\~ n}oz-Tu{\~ n}{\' o}n, C., Caon, 
N., \& Aguerri, J.~A.~L.\ 2004, \aj, 127, 58 
\bibitem[Naim \& Lahav 1997]{naim97} Naim, A.~\& Lahav, O.\ 
1997, \mnras, 286, 969
\bibitem[O'Connell et al. 1992]{o'connell} O'Connell, R.~W.~et 
al.\ 1992, \apjl, 395, L45 
\bibitem[Papovich et al. 2003]{papovich} Papovich, C., 
Giavalisco, M., Dickinson, M., Conselice, C.~J., \& Ferguson, H.~C.\ 2003, 
\apj, 598, 827 
\bibitem[Reichen et al. (1994)]{reichen} Reichen, M., Kaufman, 
M., Blecha, A., Golay, M., \& Huguenin, D.\ 1994, \aaps, 106, 523 
\bibitem[Ryder et al. 1995]{ryder} 
Ryder, S.~D., Staveley-Smith, L., Malin, D., \& Walsh, W.\ 1995, \aj, 109, 
1592 
\bibitem[Sandage 1961]{sandage61} Sandage, A.\ 1961, Washington: 
Carnegie Institution, 1961
\bibitem[Schade et al. (1995)]{schade95} Schade, D., Lilly, 
S.~J., Crampton, D., Hammer, F., Le Fevre, O., \& Tresse, L.\ 1995, \apj, 
451, L1
\bibitem[Schweizer 1980]{schweizer} Schweizer, F.\ 1980, \apj, 
237, 303 
\bibitem[Scoville (2003)]{scoville} Scoville, N.\ 2003, Journal 
of Korean Astronomical Society, 36, 167 
\bibitem[Seigar 2002]{seigar} Seigar, M.~S.\ 2002, \aap, 393, 
499 
\bibitem[S{\' e}rsic (1973)]{sersic} S{\' e}rsic, J.~L.\ 1973, 
\pasp, 85, 103 
\bibitem[Smith et al. 1996]{smith} Smith, D.~A.~et al.\ 
1996, \apjl, 473, L21 
\bibitem[Stecher et al. (1997)]{stecher} 
Stecher, T.~P., Cornett, R.~H., Greason, M.~R., ~et al.\ 1997, \pasp, 109, 584. 
\bibitem[Stoughton et al. (2002)]{stoughton} Stoughton C., Lupton R.H, Bernardini M.A., et al.\ 2002, \apj, 123, 485
\bibitem[Thornley \& Mundy  1997]{thornley} Thornley, M.~D.~\& 
Mundy, L.~G.\ 1997, \apj, 484, 202 
\bibitem[van den Bergh et al. (1996)]{van den bergh} van den Bergh, 
S., Abraham, R.~G., Ellis, R.~S., Tanvir, N.~R., Santiago, B.~X., \& 
Glazebrook, K.~G.\ 1996, \aj, 112, 359 
\bibitem[van den Bergh (2002)]{2002PASP..114..797V} van den Bergh, S.\ 2002, 
\pasp, 114, 797 
\bibitem[Waller, Gurwell, \& Tamura 1992]{waller92} Waller, 
W.~H., Gurwell, M., \& Tamura, M.\ 1992, \aj, 104, 63 
\bibitem[Waller et al. (2001)]{waller} Waller, W.~H.~et al.\ 
2001, \aj, 121, 1395 


\end{thebibliography}
\end{document}